\documentclass[twocolumn,aps,prl,amsmath,amssymb,groupedaddress]{revtex4-1}

\usepackage{graphicx}
\usepackage{graphics}
\usepackage{dcolumn}
\usepackage{bm}
\usepackage{color}
\usepackage{textcomp}
\usepackage{soul}
\usepackage{rotating}
\usepackage{setspace}
\usepackage{microtype}
\usepackage[
breaklinks=false,
bookmarks, bookmarksopen = true, bookmarksnumbered = true, linktocpage, hyperindex=true,
pdftitle={The Physical Basis for Long-lived Electronic Coherence
in Photosynthetic Light Harvesting Systems},colorlinks=true,pagebackref=false,citecolor=blue,plainpages=false,
pdfpagelabels,linkcolor=blue,urlcolor=blue]{hyperref}

\def\XXint#1#2#3{{\setbox0=\hbox{$#1{#2#3}{\int}$}
\vcenter{\hbox{$#2#3$}}\kern-.5\wd0}}

\begin{document}

\title{
The Physical Basis for Long-lived Electronic Coherence
in Photosynthetic Light Harvesting Systems}

\author{Leonardo A. Pach\'on}
\affiliation{Chemical Physics Theory Group, Department of Chemistry and
Center for Quantum Information and Quantum Control,
\\ University of Toronto, Toronto, Canada M5S 3H6}
\author{Paul Brumer}
\affiliation{Chemical Physics Theory Group, Department of Chemistry and
Center for Quantum Information and Quantum Control,
\\ University of Toronto, Toronto, Canada M5S 3H6}

\begin{abstract}
The physical basis for observed long-lived electronic coherence in
photosynthetic light-harvesting systems is identified using an  analytically
soluble model. Three physical features are found to be responsible for their
long coherence lifetimes: i) the small energy gap between excitonic states, ii) the
small ratio of the energy gap to the coupling between excitonic states, and
iii) the fact that the molecular characteristics place the system in 
an effective low temperature
regime, even at ambient conditions. Using 
this approach, we obtain
decoherence times for a dimer model with FMO parameters
of $\approx$~160~fs at 77~K and $\approx$~80~fs at
277~K. As such, significant oscillations are found to persist for
600~fs and 300~fs, respectively, in accord with the experiment and with
previous computations. Similar good agreement is found for PC645
at room temperature, with oscillations persisting for 400~fs.
 The analytic expressions obtained provide
direct insight into the parameter dependence of the decoherence
time scales. 
\end{abstract}

%
\maketitle

Electronic energy transfer is ubiquitous in nature and its
dynamics and manipulation is of special interest in diverse fields
of physics, chemistry, biology and engineering. Under natural
conditions loss of coherence is expected to occur on ultrashort
times scale due to interaction with the environment. For example,
results on betaine dye molecules \cite{HR04b} and on femtosecond
dynamics and laser control of charge transport in
trans-polyacetylene \cite{FSB08} suggest that these time scales
are very short, $\sim$2.5~fs and $\sim$3.7~fs, respectively. At
high temperatures and for weak coupling to the environment, a
classical treatment of the thermal fluctuations suggests that this
time scale can be determined as $\tau_{\mathrm{G}} =
\sqrt{\hbar^2/2\lambda k_{\mathrm{B}} T}$ where $\lambda$ is the
system reorganization energy \cite{HR04,GM08}. Based on this
expression, the dephasing time for
photosynthetic complexes, the systems of interest in
this paper, (wherein a typical value of the reorganization energy
is $\lambda = 130$~cm$^{-1}$),
can be estimated \cite{CF09} to be $\tau_{\mathrm{G}} = 45$~fs at
$T=77$~K and $\tau_{\mathrm{G}} = 23$~fs and $T=294$~K.

By contrast, recent experiments in photosynthetic complexes such
as the FMO complex \cite{EC&07,PH&10} and the PC645 complex
\cite{CW&10}  have found that electronic coherences among
different chromophores survive up to 800~fs at $77$~K \cite{EC&07}
and up 400~fs at room temperature \cite{CW&10,PH&10}. This
surprising observation and its possible consequences for
biological processes have been discussed extensively
\cite{EC&07,PH08,PH&10,CF09,RM&09,IF09,WL&10,CW&10,BE11,HC11,SR&11,NBT11} and
very elaborate models have been developed in order to understand
the underlying dynamics  \cite{IF09,BE11,HC11,SR&11,NBT11}.
Interestingly, despite the diversity of approaches and techniques,
most \cite{CF09,RM&09,IF09,CW&10,BE11,HC11,SR&11,NBT11} now
predict long-lived coherences on the same times scales as found
experimentally \cite{EC&07,CW&10,PH&10}. This suggests that the
underlying physical features are correctly contained in these
approaches. However, the sheer complexity of these computations
has limited one from identifying these essential physical
features.

In this paper we present a simple analytic approach that
provides deep insights into the long lived coherences in the
evolution of FMO complexes
\cite{EC&07,PH&10,CF09,RM&09,IF09,CW&10,BE11,SR&11} and PC645
\cite{CW&10,HC11} and allows for the identification of central
characteristics responsible for these long lived coherences. Our
analysis identifies the small \emph{effective} temperature of the
system (see below), the very small, but nonzero, energy gap
between exciton states, and their coupling as the basic elements
behind these long lifetimes. Given these conditions, we show that
the lifetimes are not ``surprisingly long''. As such, the challenge now
reverts to, for example, obtaining an atomistic understanding \cite{SR&11} of
the origins of these parameter values.

Consider first results for the Fenna-Matthews-Olson (FMO) complexes,
in particular the FMO pigment-protein complex from \emph{Chlorobium
tepidum} \cite{EC&07,PH&10}. This is a trimer consisting of
identical, weakly interacting monomers \cite{AR06}. Each weakly
interacting FMO monomer contains seven coupled bacteriochlorophyll-$a$
(BChl$a$) chromophores arranged asymmetrically, yielding seven
nondegenerate, delocalized molecular excited states (excitons)
\cite{EC&07,PH&10}. Since the electronic coupling between the BChl$a$~1
and BChl$a$~2 is relatively strong in comparison with the other
coupling strengths \cite{AR06}, we can approximate the dynamics of the
excitation as given by a dimer composed by BChl$a$~1 and BChl$a$~2. This,
often adopted, approximation is utilized below.

\begin{figure}
\begin{center}
\includegraphics[width = \columnwidth]{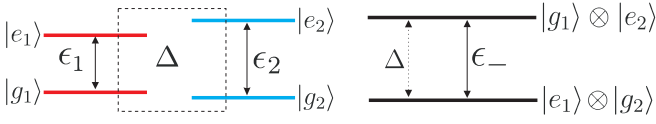}
\caption{Left hand side: The pair of interacting chromophores.
Right hand side: the
effective light harvesting two-level system
formed from the pair of
interacting chromophores \cite{GM06,ERT09}.}
\label{qubit}
\end{center}
\end{figure}

We consider the dimer (see \ref{qubit}) to be described by the
Hamiltonian \cite{GM05,GM06}
\begin{align}
\label{equ:DefInitHamilt}
\begin{split}
H &= \frac{\hbar}{2}\epsilon_1 \sigma_{z,1} + \frac{\hbar}{2}\epsilon_2 \sigma_{z,2}
  + \frac{\hbar}{2}\Delta\left(\sigma_{x,1} \sigma_{x,2} + \sigma_{y,1} \sigma_{y,2} \right)
\\
  &+\frac{\hbar}{2}\delta\mu_1\sigma_{z,1} R_1 + \frac{\hbar}{2}
\delta\mu_2\sigma_{z,2} R_2
  + B_1 + B_2,
\end{split}
\end{align}
where
$R_{i} = \sum_{\alpha}C_{\alpha,i}\left(a_{\alpha,i} + a_{\alpha,i}^{\dagger}\right)$
is the reaction field operator for molecule $i$,
$B_{i} = \sum_{\alpha}\hbar \omega_{\alpha,i} a_{\alpha,i}^{\dagger} a_{\alpha,i}$
is the energy stored in the solvent cage of molecule $i$ and $\delta\mu_j$ is
the difference between the dipole moment of the chromophore $j$ in the ground and excited
states \cite{GM05,GM06}.
The first two terms in \ref{equ:DefInitHamilt} are the
contributions from the individual sites and the third term is the $\Delta$
coupling between them. The subsequent terms describe the system-bath
coupling.
Following references \citenum{GM06} and \citenum{ERT09}, the Hamiltonian in
\ref{equ:DefInitHamilt} can be written with respect to the basis
$\{|g_1\rangle \otimes |g_2\rangle, |g_1\rangle \otimes
|e_2\rangle, |e_1\rangle \otimes |g_2\rangle,|e_1\rangle \otimes
|e_2\rangle \}$ describing the state of the two chromophores, i.e.
\begin{align}
\label{eq:H_2C}
&H=\sum_{\stackrel{\alpha}{i=1,2}}\hbar\omega_{\alpha,i}a_{\alpha,i}^{\dagger}a_{\alpha,i}
\\
&+\frac{\hbar}{2}\left(
\begin{array}{cccc}
   -\left(\epsilon_+ + V_+\right) & 0 & 0& 0   \\[2mm]\nonumber
0& -\left(\epsilon_- + V_-\right) & 2\Delta & 0\\[2mm]\nonumber
0 & 2\Delta & \epsilon_- + V_- &0\\[2mm]\nonumber
0 & 0 & 0 & \epsilon_+ + V_+
\end{array}\right)\nonumber\,,
\end{align}
where $\epsilon_{\pm}\equiv\epsilon_1\pm\epsilon_2$, and
$V_{\pm}\equiv \delta\mu_1 R_1 \pm \delta\mu_2  R_2 $.

Since under excitation by weak light only the singly excited
states need to be taken into account, we can identify
\cite{GM06,ERT09} the active environment coupled
2D-subspace as
$\{|e_1\rangle\otimes|g_2\rangle,|g_1\rangle\otimes|e_2\rangle\}$.
In this central subspace of \ref{eq:H_2C}, the effective
interacting biomolecular two-level system Hamiltonian reads
\begin{equation}
\label{eq:Baqubit}
H=\left(\frac{\hbar\epsilon}{2}\sigma_z+\hbar\Delta\sigma_x\right)+
\frac{\hbar}{2}\sigma_z V +
\sum_{\stackrel{\alpha}{i=1,2}}\hbar\omega_{\alpha,i}a_{\alpha,i}^{\dagger}a_{\alpha,i}\,
\end{equation}
where $\epsilon \equiv \epsilon_-$ and $V \equiv V_-$. This is schematically
illustrated in \ref{qubit}, where $\Delta$ is the associated ``tunneling
energy", between the new basis states $|e_1\rangle\otimes|g_2\rangle$ and
$|g_1\rangle\otimes|e_2\rangle$.

Given the biophysical nanostructure composition \cite{GM06}, we can assume that the two bath are
uncorrelated $[a_{\alpha,1},a_{\alpha,2}^{\dagger}] = 0$, which
implies that $\langle R_1(t'')R_2(t')\rangle = \langle R_2(t'')R_1(t')\rangle = 0$.
Hence, \ref{eq:Baqubit} can be written in the standard form of the spin-boson model
\cite{GM06}
\begin{equation}
H=\left(\frac{\hbar\epsilon}{2}\sigma_z+\hbar\Delta\sigma_x\right)+
\frac{\hbar}{2}\sigma_z \sum_{\beta}g_{\beta}(b_{\beta} + b_{\beta}^{\dagger}) +
\sum_{\beta}\hbar\omega_{\beta}b_{\beta}^{\dagger}b_{\beta}\ ,
\label{eq:Bqubit}
\end{equation}
where the $b_{\beta}$ includes harmonic oscillators coupled to both
chromophores, with couplings $g_{\beta}$.

The environment is fully characterized by the spectral density
$J(\omega)=\sum_{\alpha}g_{\alpha}^2\delta(\omega-\omega_{\alpha})$,
being a quasi-continuous function for typical condensed phase
applications that determines all bath-correlations that are
relevant for the system \cite{Wei08,LC&87}.
For Ohmic dissipation $J(\omega) = 2 K \omega
\,\mathrm{e}^{-\omega/\omega_{\mathrm{C}}}$, where the
dimensionless parameter $K$ describes the
damping strength and $\omega_{\mathrm{C}}$ is the cut-off frequency
\cite{LC&87,Wei08}. An Ohmic spectral density is a useful choice
for, e.g. electron transfer dynamics or biomolecular complexes
\cite{MK01,GM06}. The parameter $K$ is related to the
reorganization energy $\lambda$ by means of $\lambda = 2 K \hbar
\omega_{\mathrm{C}}$ and the phonon relaxation time is given by
$\tau = \pi/(2 \omega_{\mathrm{C}})$ \cite{Wei08,IF09}.

The Hamiltonian in \ref{eq:Bqubit} has been extensively studied in
literature (cf. Chaps.~18-22 in Ref.~\citenum{Wei08} and references
therein). The parameter range within which the light harvesting systems
of interest lie allows for the use of the non-Markovian
non-interacting blip approximation (NIBA) plus first order 
corrections  in the interblip correlation strength, i.e. 
an {\it enhanced NIBA approximation}. This approximation is
valid for weak system-bath coupling
and for $\epsilon/2\Delta < 1$, over the whole range of temperatures (see
Chap.~21 in Ref.~\citenum{Wei08}), and provides simple and accurate
analytic expressions for relaxation and decoherence rates.

In the case of FMO, the energy gap $\epsilon$ is $(315 -
240)$~cm$^{-1}$ = 75~cm$^{-1}$ while the coupling energy $\Delta$ corresponds to
87.7~cm$^{-1}$ (see \cite{IF09,NBT11} and references therein).
In accord with 
references \citenum{IF09} and \citenum{NBT11}, the reorganization energy in this
case is $\lambda \approx 35$~cm$^{-1}$ and the phonon
relaxation time is $\tau=50$~fs. Hence,  for this case we get
$K=\lambda/(2\hbar\omega_{\mathrm{C}}) = \lambda \tau/(\hbar \pi) =  0.105$.
Additionally, at $T = 77$~K, $2\Delta/k_{\mathrm{B}} T \approx
3.28$ while at $T = 277$~K, $ 2\Delta/k_{\mathrm{B}} T \approx
0.911$. \ref{tab:parametersFMO} summarizes the parameters of
the present analysis, where we have used the same $\omega_{\mathrm{C}}$
value at both temperatures. Clearly these parameters place the system within
the domain of accuracy of the enhanced  
NIBA approach. We emphasize that this selection of 
parameters is widely used in the literature \cite{IF09,NBT11}, 
and it is not chosen to fit our model to the experimental results.

\begin{table}[h!]
\begin{center}
\begin{tabular}{c|c|c|c}\hline
 $\epsilon/2\Delta$ & $K$ & $2\Delta/\omega_{\mathrm{C}}$ &
$2\Delta/k_{\mathrm{B}} T$
\\
\hline \hline 0.428 & 0.105& 1.052 & 3.28 \\ \hline
0.428 & 0.105& 1.052 & 0.911 \\ \hline
\end{tabular}
\end{center}
\caption{Parameters used for dimer formed of BChl$a$~1 and BChl$a$~2 at $T=77$~K (first row) and
$T=277$~K (second row).}
\label{tab:parametersFMO}
\end{table}

The high temperature limit in this approach is given by
temperatures well in excess of
$T_{\mathrm{b}} = \hbar (\Delta_{\mathrm{eff}}^2 + \epsilon^2)^{1/2}/k_{\mathrm{B}}$, where
\begin{equation*}
\Delta_{\mathrm{eff}} =
[\Gamma(1-2K)\cos(\pi K)]^{1/2(1-K)}(\tilde{\Delta}/\omega_{\mathrm{c}})^{K/(1-K)}\tilde{\Delta},
\end{equation*}
where in our case $\tilde{\Delta} = 2 \Delta$.

For the set of parameters listed in \ref{tab:parametersFMO},
we find $T_{\mathrm{b}} \approx 288$~K. Hence the FMO experiments,
at 77 K and 277 K, are in the low temperature regime, $T <
T_{\mathrm{b}}$. In this regime, the Rabi
frequency $\Omega$, the relaxation rate $\gamma_{\mathrm{r}}$ and
the decoherence rate $\gamma$ are given by \cite{Wei08}
\begin{align}
\label{equ:OmegaG}
\Omega^2 &= \Delta_{\mathrm{eff}}^2[1-2\Re u(\mathrm{i}\Delta_{\mathrm{b}})] + \epsilon^2
\\
\label{equ:gammarG}
\gamma_{\mathrm{r}} &= \frac{\pi}{2} \frac{\Delta_{\mathrm{eff}}^2}{\Delta_{\mathrm{b}}^2}
S(\Delta_{\mathrm{b}})
\\
\label{equ:gammaG}
\gamma &= \frac{\gamma_{\mathrm{r}}}{2} + \frac{\pi}{2} K \frac{\epsilon^2}{\Delta_{\mathrm{b}}^2}S(0),
\end{align}
respectively, where
$\Delta_{\mathrm{b}} = \sqrt{\Delta_{\mathrm{eff}}^2 + \epsilon^2}$ and
$ u(z) = \frac{1}{2}\int_0^{\infty} \mathrm{d} \omega \frac{J(\omega)}{\omega^2 + z^2}
(\coth(\hbar \omega/2k_{\mathrm{B}}T)-1)$
while $S(\omega) = J(\omega) \coth(\hbar \omega/2k_{\mathrm{B}}T)$
is the noise power. Non-Markovian corrections are already included
in \ref{equ:OmegaG} and \ref{equ:gammaG}.
For the particular case of Ohmic dissipation
adopted here \cite{Wei08},
\begin{align}
\begin{split}
\label{equ:Omega}
\Omega^2 &= \Delta_{\mathrm{b}}^2
\\ & + 2 K \Delta_{\mathrm{eff}}^2
           \left[\Re\psi(\mathrm{i}\hbar\Delta_{\mathrm{b}}/2\pi k_{\mathrm{B}} T)
     - \ln(\hbar\Delta_{\mathrm{b}}/2\pi k_{\mathrm{B}} T)\right]
\end{split}
\\
\label{equ:gammar}
\gamma_{\mathrm{r}} &= \pi K \coth(\hbar \Delta_{\mathrm{b}}/2k_{\mathrm{B}}T)
\Delta_{\mathrm{eff}}^2/\Delta_{\mathrm{b}},
\\
\label{equ:gamma}
\gamma &= \gamma_{\mathrm{r}}/2 + 2 \pi K (\epsilon^2/\Delta_{\mathrm{b}}^2)k_{\mathrm{B}} T/\hbar,
\end{align}
where $\psi(z)$ is the digamma function. \ref{equ:gammar} - \ref{equ:gamma} provide simple analytic
expressions for the desired rates.

\begin{figure}
\includegraphics[width = \columnwidth]{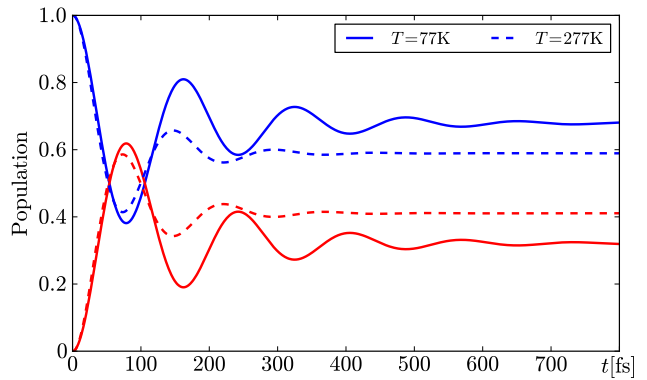}
\caption{Time evolution of the population of the electronic
manifolds $|e_1\rangle \otimes |g_2\rangle \langle g_2 | \otimes
\langle e_1 |$ (blue curves) and $|g_1\rangle \otimes |e_2\rangle
\langle e_2 | \otimes \langle g_1 |$ (red curves) for the FMO dimer
formed of BChl$a$~1 and BChl$a$~2 using the parameters displayed
in \ref{tab:parametersFMO}. Continuous curves depict the
evolution at $T=77$~K whereas dashed curves are at $T=277$~K. We
assume that at $t=0$, the full excitation is localized in
$|e_1\rangle \otimes |g_2\rangle \langle g_2 | \otimes \langle e_1
|$.} \label{fig:populationsFMO}
\end{figure}

With this set of expressions, and the parameters given in
\ref{tab:parametersFMO}, we find $2 \pi \Omega^{-1} =
163$~fs, $\gamma_{\mathrm{r}}^{-1} = 90$~fs, $\gamma^{-1} =
153$~fs at $T=77$~K, while
$2 \pi \Omega^{-1} = 151$~fs, $\gamma_{\mathrm{r}}^{-1} = 45$~fs,
$\gamma^{-1} = 69$~fs at $T=277$~K.
Despite the simplicity of the model, the
resultant dynamics [given analytically, for general spectral densities,
in Eqs. (21.79), (21.170) and (21.173)
of Ref.~\citenum{Wei08}] is depicted in \ref{fig:populationsFMO} and
describes the survival of coherences on the correct time scale and
in good agreement with  recent results\cite{IF09,SR&11,NBT11}. 
In \ref{fig:populationsFMO}, only the
decay of the excitations is absent, since coupling to other
chromophores \cite{IF09,SR&11,NBT11} is here neglected.
The global decay in \ref{fig:populationsFMO} is a 
result of two processes: one associated with the 
$\gamma_{\mathrm{r}}$- relaxation of the
dynamics, and a second one related to the 
loss of coherence due to the presence of the thermal bath, associated with 
$\gamma$.

\begin{figure}
\includegraphics[width = \columnwidth]{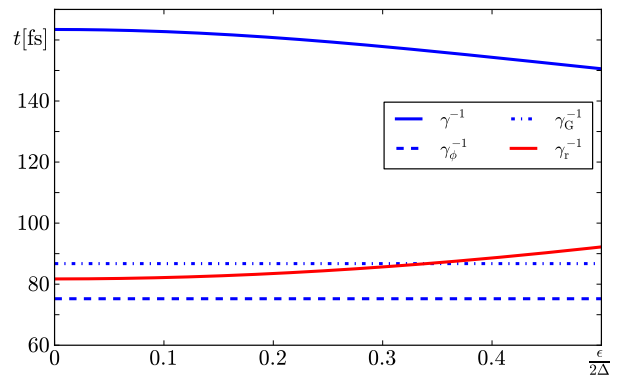}
\caption{Relaxation time $\gamma_{\mathrm{r}}^{-1}$ (red line) and
decoherence time $\gamma^{-1}$ (continuous blue line) as functions
of $\epsilon/2\Delta$, for fixed $\Delta= 35$ cm$^{-1}$ and 
at $T=77$~K.
The dashed blue line, $\gamma_{\phi} =
2\pi (k_{\mathrm{B}} T/\hbar)\lambda/\hbar \omega_{\mathrm{C}}$,
denotes the dephasing rate used in references \citenum{GM08,RM&09,PH&10} and the
dotdashed blue line denotes the decoherence time
$\tau_{\mathrm{G}} = \sqrt{\hbar^2/2\lambda k_{\mathrm{B}} T}$
used in references \citenum{HR04,GM08,CF09}. Fixed parameters as in
\ref{tab:parametersFMO}.} \label{gammaVSepsilon}
\end{figure}

To examine the parameter dependence, \ref{gammaVSepsilon}
shows the relaxation rate $\gamma_{\mathrm{r}}$ and the
decoherence rate $\gamma$ as a function of the energy splitting,
$\epsilon$, for fixed $\Delta$, using the parameters displayed in
\ref{tab:parametersFMO}. Clearly in this ``low temperature"
regime, i.e., $T < T_b$, the
larger the ratio $\epsilon/2\Delta$ the shorter the dephasing
time and concomitantly the longer the relaxation time. These time
scales are also compared in \ref{gammaVSepsilon}
with the ones predicted by $\gamma_{\phi}
= 2\pi (k_{\mathrm{B}} T/\hbar)\lambda/\hbar \omega_{\mathrm{C}}$
used in references \citenum{GM08,RM&09,PH&10} and with $\gamma_{\mathrm{G}}= 1/\tau_{\mathrm{G}}$
discussed earlier, both of which are clearly far too short.

As a second example we consider results on marine algae
\cite{CF09}, in particular  the results for PC645, which has been
studied experimentally \cite{CW&10},
and numerically in great detail \cite{HC11}. PC645 contains
eight bilin molecules covalently bound to the protein scaffold. A
dihydrobiliverdin (DBV) dimer is located at the center of the
complex and two mesobiliverdin (MBV) molecules located near the
protein periphery give rise to the upper half of the complex
absorption spectrum. Excitation of this dimer initiates the light
harvesting process \cite{CW&10}. The electronic coupling between
the closely spaced DBVc and DBVd molecules is $\sim$320 cm$^{-1}$
and this relatively strong coupling results in delocalization of
the excitation, yielding the dimer electronic excited states
labelled DBV$+$ and DBV$-$. Excitation energy absorbed by the
dimer flows to the MBV molecules which are each 23 {\AA} from the
closest DBV, and ultimately to four phycocyanobilins (PCB) that
absorb in the lower-energy half of the absorption spectrum
\cite{CW&10}.

The exciton states related to DBVc and DBVd are mainly
composed of DBV$-$ and DBV$+$ that are antisymmetric and symmetric
linear combinations of the DBV sites, though they also contain
small contributions from the other bilin sites \cite{HC11}. This
allows us to concentrate only on a dimer, as in the previous
example,  here formed of DBVc and DBVd. In this case, a
Debye-Ohmic spectral density, $J(\omega) = 2 \lambda \omega
\tau/(1+\omega^2\tau^2)$, is more appropriate \cite{HC11}.

\begin{figure}
\includegraphics[width = \columnwidth]{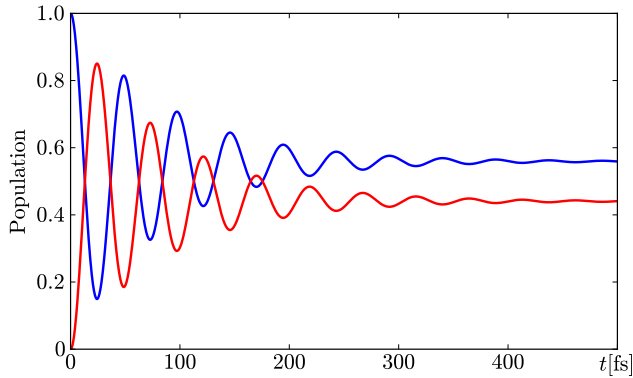}
\caption{Time evolution of the population of the electronic manifolds
$|e_1\rangle \otimes |g_2\rangle \langle g_2 | \otimes \langle e_1 |$ (blue curve) and
$|g_1\rangle \otimes |e_2\rangle \langle e_2 | \otimes \langle g_1 |$ (red curve) for the
PC645 DVB-dimer.}
\label{fig:populationsPC645}
\end{figure}

For the chromophores DVBc and DVBd the energy gap is
$(17116 - 17034)$~cm$^{-1}$ = 82~cm$^{-1}$ with the coupling
energy corresponding to 319.4~cm$^{-1}$ \cite{CW&10,HC11}.
In accord with Ref.~\citenum{HC11}, the reorganization energy $\lambda$ in
this case is $\sim$130~cm$^{-1}$, with the shorter of two relaxation
times being $\tau=50$~fs. Additionally,
at $T = 294$~K, the temperature at which the experiment was
performed, $2\Delta / k_{\mathrm{B}} T = 3.1262$.
For this set of parameters, we have $T_{\mathrm{b}} \sim 926$~K, a
high temperature that is consistent with the high frequencies
involved in the present case \cite{GPZ10}. Hence the PC645
experiment, at room temperature, is also in the  effective
low temperature $T < T_{\mathrm{b}}$ domain. For the case of the
Debye-Ohmic spectral density we have (via a relatively simple numerical
computation) that
$2 \pi \Omega^{-1} =
49$~fs, $\gamma_{\mathrm{r}}^{-1} = 76$~fs, $\gamma^{-1} = 88$~fs with
the associated evolution shown in
\ref{fig:populationsPC645}. These results are in very good qualitative
agreement with both the long lived experimental coherence time scales and
with the recent intricate  solution to the master
equations \cite{HC11}.

Considering that the long time scales emerge naturally here from the
system parameters, why were far shorter decoherence time scales
originally expected for these systems?  To see this, note that in
molecular systems, dynamics  is often studied between different
electronic eigenstates of the system, separated by greater than
$\sim 10^{4}$~cm$^{-1}$, with  no coupling between them.
In such cases, the dephasing time from \ref{equ:gammar}
and \ref{equ:gamma} would be extremely short. By contrast, in the
case of photosynthetic complexes, energy transfer occurs between
exciton states that are close in energy and additionally are
coupled.  This generates a small value for the ratio
$\epsilon/\Delta$ which in turn is responsible for longer
dephasing times (see \ref{gammaVSepsilon}). Additionally,
expression such as $\tau_{\mathrm{G}}$, which are often
used to estimate rates, are only valid at high temperature,
$k_{\mathrm{B}} T \gg \hbar \omega_{\mathrm{C}}$, and at short
times, $t < \omega_{\mathrm{C}}^{-1}$. Under conditions when the expression for
$\tau_\mathrm{G}$ is valid, the bath modes can be treated
classically \cite{TWM01}, as in references \citenum{HR04b} and \citenum{FSB08}. When one is not in
the appropriate regime,  the
classical evolution of the bath underestimates quantum coherence
effects \cite{TWM01} because at low temperatures quantum
fluctuations overcome thermal fluctuations \cite{HR04}. Hence, estimates
based on $\tau_G$ are unreliable. Similarly,
$\gamma_{\phi} =
2\pi (k_{\mathrm{B}} T/\hbar)\lambda/\hbar \omega_{\mathrm{C}}$ estimates
also provide an inadequate representation of the true physics and
associated dependence on system and bath parameters, and result in
decoherence times that are severely underestimated and
misleading.

In summary, a proper spin-boson treatment of electronic energy transfer in model
photosynthetic light harvesting systems has been shown to
give {\it analytic} results with long
coherence lifetimes that are in very good  agreement with experiment and
with other, far far more complex, computations. The analytic form allows an
analysis of the parameter dependence of the decoherence times, and
shows that the observed long lifetimes arise naturally in the
effective low temperature regime and for appropriate
ratios of the energy splitting to the coupling strength 
and are, in this sense, not surprisingly long.  
Further, the model has predictive power, allowing one to identify other parameter
ranges over which longlived coherences will exist. 

A more detailed analysis will involved
extending this model to the 7 or 8-site system and to second order correction in the interblip
approximation; work in this direction is in progress. Finally, in addition to
examining other light harvesting systems, the possible 
application of this approach to
other systems displaying long-lived coherences, such as superpositions
of two excitons in quantum dots \cite{HK&07} is of interest.

\acknowledgements
This work was supported by the US Air Force Office of Scientific Research under contract number
FA9550-10-1-0260.

\bibliography{llcprl}

\end{document}